# Together we stand, Together we fall, Together we win: Dynamic Team Formation in Massive Open Online Courses

Tanmay Sinha
School of Computer Science
Vellore Institute of Technology
Chennai 600127, India
tanmay.sinha655@gmail.com

*Abstract*—**Massive Open Online Courses (MOOCs) offer a new scalable paradigm for e-learning by providing students with global exposure and opportunities for connecting and interacting with millions of people all around the world. Very often, students work as teams to effectively accomplish course related tasks. However, due to lack of face to face interaction, it becomes difficult for MOOC students to collaborate. Additionally, the instructor also faces challenges in manually organizing students into teams because students flock to these MOOCs in huge numbers. Thus, the proposed research is aimed at developing a robust methodology for dynamic team formation in MOOCs, the theoretical framework for which is grounded at the confluence of organizational team theory, social network analysis and machine learning. A prerequisite for such an undertaking is that we understand the fact that, each and every informal tie established among students offers the opportunities to influence and be influenced. Therefore, we aim to extract value from the inherent connectedness of students in the MOOC. These connections carry with them radical implications for the way students understand each other in the networked learning community. Our approach will enable course instructors to automatically group students in teams that have fairly balanced social connections with their peers, well defined in terms of appropriately selected qualitative and quantitative network metrics.**

## I. Introduction

As Massive Open Online Courses(MOOCs) continue to proliferate in the realm of online education, we expect such a form of lifelong learning holding tremendous potential, to provide students with cognitive surplus beyond traditional forms of tutelage. With scaled up class sizes, there is an increased need for collaborative work to accomplish course assignments, solve queries of peer classmates to facilitate effective interaction on discussion forums, and comprehend multiple forms of computer mediated inputs from the instructor that come as video lectures, hangouts, e-books and additional study materials on the web. However, there are certain bottlenecks which question the collaborative output of MOOCs and constrain benefits such as work efficiency, students' bonding with classmates, accountability and learning opportunities, that arise from teamwork. Collaborations become difficult in the computer mediated MOOC environment, because students are mostly geographically distributed. With students flocking to these MOOCs from a motley of backgrounds and participating without face to face communication, it becomes difficult for an instructor to manually organize the students into study groups which can work together as teams, towards fruitfully achieving course objectives. Furthermore, the instructors' perception of students' cognitive level when working as a team, might not be a true reflector of their actual learning. This can adversely affect the instructor's evaluation of group dynamics (goal setting, progress, knowledge building), and he can face difficulty in perceiving when to provide support [20]. Ultimately, it is the students who are at loss, because of issues such as unsatisfactory learning outcomes, coordination problems etc.

Automatic team formation can thus have a substantial impact on MOOCs, by improving the overall learning process, making knowledge transfer more fluidic, and alleviating students' concerns regarding (i)whom to seek help from (one of the potential reasons which could be attributed to huge attrition rates); (ii)provision of a channel for increasing two way interactions and extended discussions. This will also boost students' problem solving skills, by guiding them to engage in close discussion with suggested teammates. Because teams provide a lens into the behavior and social system of MOOCs as a whole, it is more feasible for a course instructor to monitor interactions in MOOCs from this structured perspective, rather than haphazardly tracking engagement patterns of individual students.

Once we dynamically form teams, it opens up a plethora of interesting research directions including investigation of how these teams collaborate to mutual agreement(advantage) or conflict(disadvantage), influential factors that distinguish good teams from the bad teams, and how we could leverage characteristics of good teams for assisting educational designers to establish a pedagogical basis for decision-making while designing MOOCs. There is an impressive body of literature that backs up the study of important variables in organizational team study such as effectiveness and productivity [1], effect of diversity on team functioning [2], role of conflict among teammates [3], leadership, motivation and creativity [4]. Thus, understanding the impact of better team formation methodologies on team performance, collective efficacy and group cohesion in MOOCs is a fascinating research direction. The scientific challenge is to zoom into the pattern of connections among students in MOOCs, to better understand the relation between team formation methodologies and collaborative work output.

In this paper, we first describe our motivation in Section II. Then, the crisp objective is outlined in Section III. In Section IV, we describe our proposed approach for dynamic team formation in MOOCs in detail. Section V deals with the literature survey. We talk about dataset description and future work in Section VI. We end with conclusion in Section VII.

## II. MOTIVATION

With rapid surge in development of MOOCs, Kizilcec et al. [14] have stressed on the importance of providing guidance to MOOC learners for group collaboration. Using evidence from 21 MOOCs offered within a time span of 2 years, the feasibility of geographically distributed groups (which rely only on computer mediated communication) and in-person groups (which additionally rely on face to face interaction, because of lying in close physical proximity and being based on the idea of homophily) was outlined. The sudden rise in interest in developing strategies to bolster group learning in MOOCs is axiomatic from the fact that, researchers at Stanford and Carnegie Mellon (pioneers of open learning initiatives) have also started working on ideas that focus on the development of algorithms for automatic team creation and study of team interactions to understand team members' engagement. However, prominent MOOC offerings like Coursera, Udacity, EdX etc still lack such a mechanism design for team creation and effectiveness evaluation.

We can represent MOOCs as a social network, with nodes representing students and edges denoting the communication patterns among them. Despite the applicability of traditional social network techniques to the MOOC space, it is important to understand that community discovery in social networks is a different problem from team formation in a networked learning community such as a MOOC. Prior work has intuitively defined community as a collection of nodes (individuals) that have greater ties internally, than to the rest of the network. To formalize this notion, quality metrics such as normalized cuts, conductance, modularity etc are used, while popular community detection algorithms such as Kernighan-Lin (KL), Agglomerative, Spectral, Markov Clustering and Multi level graph partitioning explicitly try to optimize these metrics. In simple terms, these metrics produce communities that are well connected amongst themselves but are sparsely connected to the rest of the graph, have low inter cluster edge weights and are fairly balanced in sizes [5].

However, in context of team formation in the MOOC network, some or all these constraints might be relaxed. Team size and membership could vary depending on task size, and we would want teams not only to effectively communicate within themselves, but also incorporate diverse perspectives for problem solving from other neighboring teams. This provides motivation towards a new formulation of the team formation problem in MOOCs.

## III. OBJECTIVE

The objective is to capitalize on the rich informal connections among students in MOOCs to develop novel ways of dynamic team formation. Additionally, we want to investigate the diverse nature of communication flows that can occur in MOOCs, and how such forms of information diffusion can potentially influence students' adaptive participation strategies in coursework.

## IV. PROPOSED APPROACH

Figure 1 summarizes our systematic approach for dynamic team formation in MOOCs. The basic aim is to:

1. Find groups of students in the MOOC who can collectively perform tasks (assignment, recitation, project etc) in an efficient manner.
2. Decipher how effectively does the information required for teamwork, diffuse along the edges of the MOOC network.

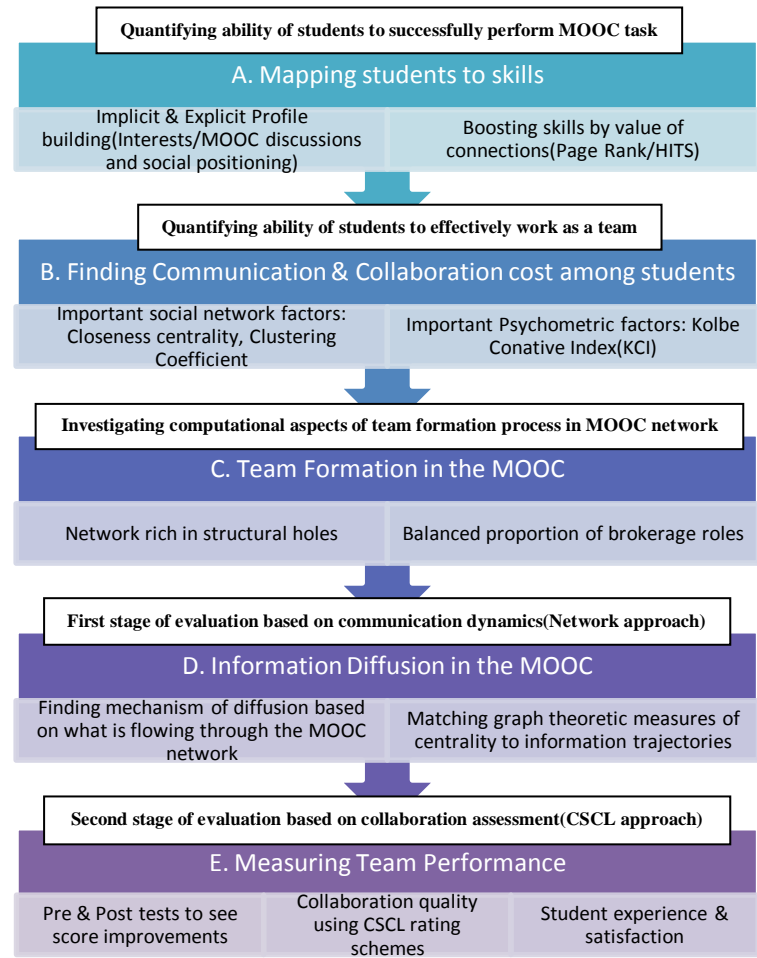

Figure 1. Summarized Steps for dynamic team formation in MOOCs

Steps A and B could be processed in parallel. The details are outlined below:

A. Associating each student with a set of skills:
   i. Following a user centric approach, we can initially use explicit profile information such as interests, while implicit skill information can be inferred from students' conversations and their social positioning within the discussions later. Specifically, a probability distribution for domain specific terms used in the MOOC context can be drawn to get the top 'k' skills for each student.
   ii. Considering the fact that students' skills can be potentially boosted by the value of their connections, graph ranking algorithms such as Page Rank and HITS(Hypertext Induced Topic selection) [24] can be applied to further refine and update the skills. These algorithms quantitatively compute the authority and hub scores for each student in the post reply networks extracted from discussion forums. In the MOOC context, students who interact with many students in discussions will have high authority scores (information content), while those who get engaged in others' initiated discussions will have high hub scores (quality of nodes' links).

B. Finding out the communication and collaboration cost (overhead incurred when team members work together) between students implicitly using factors from social network analysis:
   i. Students having low closeness centrality(sum of geodesic graph theoretic distances from other students) or high clustering coefficient (embeddedness of students in their neighborhood) will be better positioned to easily link with others for producing collaborative output. Formally, clustering coefficient $C_i$ for a vertex $v_i$ in a directed network is given by the proportion of links between the vertices within its neighborhood divided by the number of links that could possibly exist between them. In a graph $G(V, E)$, if an edge $e_{ij}$ connects vertex $v_i$ with $v_j$, $N_i$ represents the immediately connected neighbors of $v_i$ and $k_i$ represents the number of neighbors of the vertex $v_i$, $C_i=|\{e_{jk}: v_j, v_k \in N_i, e_{jk} \in E\}|/k_i(k_i-1)$. Thus, we could first examine the effect of rewiring the MOOC network to maximize clustering coefficient and minimize geodesic distances. Such an arrangement would give the optimal collaboration cost that could be used for expressing students' opportunities or constraints in working along with others in the MOOC.
   ii. Similarly, certain psychometric measures such as Kolbe Conative Index (KCI) [21] can also aid in understanding personality of team members in the MOOC, specifically about how students instinctively approach problem solving, arrange ideas or objects, use time or energy and demonstrate goal directed action.

C. Forming teams such that the following two criteria are met:
   i. Students in the teams have a network that is rich in structural holes [22]. In MOOCs, creativity will be stifled and opportunities restricted, if team members tend to focus their activities only inside their own teams. Thus, if a MOOC team comprises of students who can mobilize social capital by acting as a brokers of information between different student clusters that otherwise would not have been in contact, the team can have inundated access to new ideas, opinions and opportunities. In particular, we shall have to consider forming teams where a student (ego) has less redundant connections (alters), and these connections do not constrain or induce a dependency on the ego directly or indirectly. An example of structural hole is presented in Figure 2.
   ii. Teams have a balanced proportion of students playing the brokerage roles of coordinator, consultant, gatekeeper, representative and liaison [23]. These roles are described in Figure 3. Specifically, for finding out the five different kind of brokerages specified, we would have to examine every instance where a student lies on a directed path between other students. Using a partition of students into groups based on criteria relevant to group processes like interpersonal dynamics, demographic information, skills etc, we could count the number of times each student plays these roles.

D. Studying information diffusion models for the MOOC: It involves 2 steps:
   i. Firstly, we need to decide whether diffusion happens via parallel replication (e-mail broadcast by MOOC instructors, students' influential attitude that helps others to actively engage in MOOC discussions), serial replication (informational/emotional support provided to MOOC students) or transfer mechanism (study materials being passed among students lying in close or distant geographical proximity)
   ii. Secondly, we need to match the appropriate graph theoretic measures of centrality to trajectories (geodesics/paths/trails/walks) along which the information would flow. Though a closely related simulation was done by Borgatti et al. [19], only flows having a source and target were considered. However, in context of MOOCs, we might have cases where, though the flows originate systematically, they have no particular target.

E. Measuring outcomes for team performance: This can be done by dividing students in the MOOC into experimental and control groups and:

i. Using pre and post tests to demonstrate effectiveness and improvement in test scores after using the proposed methodology for dynamically forming teams in the MOOC.

ii. Capturing collaboration quality using Computer supported collaborative learning (CSCL) rating schemes such as Meier's rating scheme [15]. This scheme comprises of nine dimensions that cover the essence of the important aspects of collaboration such as communication, joint information processing, coordination, interpersonal relationship and individual motivation. It would be interesting to see how collaboration quality differs in informal MOOC interactions, as compared to direct communication methods such as video conferencing and face to face meetings.

iii. Implicitly inferring member experience and satisfaction by the sentiment analysis of their conversation, while explicitly doing the same through a questionnaire. Formally, an expressed opinion is defined as a quintuple ($e_j$, $a_{jk}$, $so_{ijkl}$, $h_i$, $t_l$), where $e_j$ is a target entity, $a_{jk}$ is an aspect/feature of entity $e_j$, $so_{ijkl}$ is the sentiment value of the opinion from the opinion holder $h_i$ on aspect $a_{jk}$ of entity $e_j$ at time $t_l$, $h_i$ is an opinion holder and $t_l$ is the time when the opinion is expressed [25].

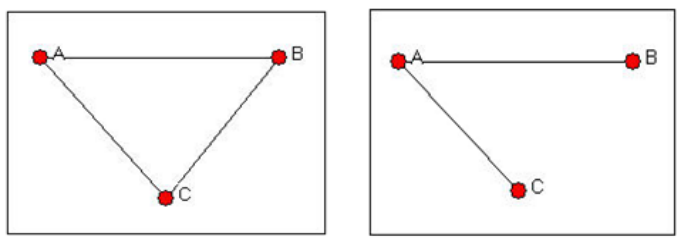

Figure 2. The left half depicts 3 students A, B, C forming a triad with no structural holes. The right half represents the same 3 actor network, but with a structural hole between student B and C [27]

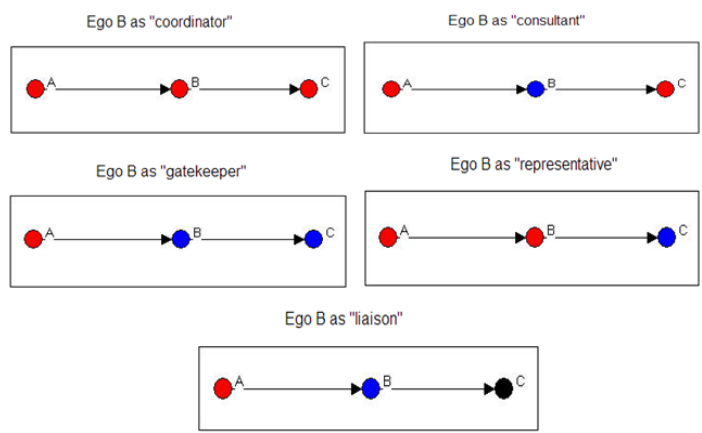

Figure 3. 5 different kinds of brokerage roles that can occur in a social network. Colored nodes represent students belonging to different groups [27]

## V. RELATED WORK

The rationale for the proposal is grounded in the theory of social network analytics(SNA), teamwork and computer mediated collaborative learning. For clarity, the related work can be divided into 2 parts: (i)SNA and team formation; (ii)SNA and team performance.

*SNA and team formation:* The alignment and integration of network theory and team theory was first done by the seminal works of David et al. [6] and to some extent by Borgatti et al. [7]. In these papers, fundamental building blocks that characterize informal interaction patterns in social networks, such as strong and weak ties(connections), social capital, centralized and decentralized networks, social cognition and embeddedness of individuals and teams, were mapped to the team literature characterized by input(team composition, nature of task), process(task and social interactions among teammates) and output(quality of team's product and viability of the team as a functioning unit). Some existing intersection between the two bodies of literature was clarified by highlighting how functional needs of teams such as knowledge transfer and coordination(simple/complex), free riding and external information needs were related to the network needs. The proposed typology of research focused on network consequences from topological(structural) as well as flow(diffusion) perspectives.

Another influential work that fueled interest in dynamic team formation was the work on multi-agent networks by Gaston et al. [8]. Based on different states of the agents such as uncommitted, committed or active, they could initiate team formation or join the team. The underlying network was dynamic because agents could adapt their network connectivity based on preferential attachment or rewire their neighborhood based on task performance. The conducted simulations showed that these team adaptation strategies heavily increased team performance, and corroborated prior evidences of team formation processes being enhanced by network topology with short average path length between nodes and presence of hub structures. In their extended set of experiments, Gaston et al. [9] further showed that some network structures were more suited to team formation than others. The interaction topology of agents in lattice, random, small-world, scale-free and networks could drastically affect their ability to cooperatively work in teams. Diversity support or the percentage of all possible skill combinations supported by the agent organization, was outlined as one of the positive factors affecting team formation, while scale free networks led to most efficient team formation.

Other prior works have studied online team formation with multi ended objectives of minimizing coordination cost, balancing workload or both. Graph theoretic concepts such as diameter (largest shortest path in a graph), Steiner trees (minimal spanning tree with a relaxed constraint that extra intermediate vertices and edges may be added to the graph in order to reduce the length of the spanning tree) have been applied to balance the tradeoff in the converted optimization problem [10][11]. Team formation has also been formulated as an integer linear programming problem for study by the Operation Research community. Standard combinatorial optimization methods such as branch and bound [16], statistical approaches such as taguchi's parameter design [17] have been applied, with a primary focus on matching experts' skills with task requirements. Ignored organizational social bonds were taken into account in team-formation methodologies by Wi et al. [18], where a set of social-network measures for identifying the effectiveness of a team, and a genetic algorithm for finding a good team of experts was proposed. Fuzzy logic was used to calculate the personal knowledge score as well as familiarity scores between employees. The mathematical model provided flexibility to adjust these factors.

*SNA and team performance:* An interesting work by Balkundi et al. [12] used meta-analytic techniques to support their hypothesis about the impact of social network properties on team performance and viability, based on factors such as density of ties expressing formal relationships(instrumental), friendships(expressive), team leader centrality and team centrality in the intergroup network. All variables concurred with the propositions, except that expressive ties turned out to be a stronger predictor of team performance than instrumental ties. The results also showed that, as team members became more familiar with each other, impact of integrative social structures on team task performance weakened. An inherent weakness in this study was that individuals' attributes were overlooked, and the focus was only on interactions among team members.

Although most of the generic concepts and mappings delineated above are applicable to teams functioning in educational scenarios also, majority of these ideas have only been applied to managerial settings, sensor networks, disaster and emergency response networks, movie databases like IMDB and social bookmarking datasets like Bibsonomy. Online education, specifically MOOCs present a new challenge for discovering effective ways of team formation, based on the underlying social network structure. Knowledge management and measures of team effectiveness can be considered more subjective in MOOCs, because there is a lot of in uncertainty in discussions among novice, intermediate and expert students. Also, there is no definitive point where knowledge building with the subject matter can terminate.

The sub-stream of research that comes close to examining the effect of correlating social network structure with online course performance of teams, is the one related to collaborative innovation networks(COIN). Gloor et al. [13] provided insights into digital collaboration dynamics by developing an online course wherein student teams had to study structural properties of various online communities. Team formation was left to students. The important point was that, they could know their social position and contribution patterns, while forming alliances. The results based on student peer to peer evaluation indicated that, teams which exhibited balanced communication behavior(equal contribution in terms of messages sent and received) performed best. However, this online collaboration task was limited to less than 50 students who were drawn from only two geographical locales. Therefore providing autonomy to students to form teams and monitoring their social network position might have been fine. But, considering the scale of MOOCs, the diversity in experience and skill levels of students participating, such simplistic measures of team formation and evaluation might not serve the purpose. This is the core problem that is being addressed by the proposal.

## VI. FUTURE WORK

In preparation for engaging in a partnership with an instructor team for a Coursera MOOC that was launched in Fall of 2013, we were given permission by Coursera to scrape and study a small number of other courses. Coursera started in April 2012 with just 30 online courses, and has now developed into a fully fledged online education portal featuring 395 courses from 84 renowned Universities extending beyond one discipline, and 9.5 million enrollments from students representing 195 countries. Our goal was to gain insights that would enable us to develop tools for instructor support. We began by informally examining the interactions on discussion threads in a literature course offered in June 2013. We collected 1503 posts and 1100 comments among 665 threads posted to the discussion forums during the first seven weeks of the course. A total of 771 students participated in the forum discussion during the seven-week period, not counting those students posting anonymously. Using the reply structure of students in the discussion forums, our constructed social network graph from interactions within the discussion forums contained a total of 3848 edges. Our prior work on dropout analysis on MOOCs [26] has considered similar dataset for experimentation. For future work, we want to apply our approach for dynamic team formation to Coursera MOOC that we have been able to study and extract data from. Generally, Coursera discussion portals are divided into forums and subforums and each of them have certain number of threads. Each thread is initiated with a "thread starter" post that serves as a prompt for discussion. The thread builds up as people start follow up discussions by their posts and comments. Thus, our data contains all posts and comments for the given period, annotated with thread, forum, subforum, author (when not posted anonymously), parent post (for comments), and timestamp.

## VII. CONCLUSION

In this work, we have outlined the perspectives and logical steps for dynamic team formation in MOOCs. Apart from considering only students' attributes, we also incorporate a fundamental aspect of humanity: students' connections. An informal social network is always there, hidden behind the MOOC, exerting both subtle and dramatic influence over students' choices, actions, thoughts, feelings, even desires. These ties, and the particular pattern of these ties, are often more important than the individual students themselves. They allow groups of students to do things that a disconnected collection of individuals cannot. The ties explain why the whole is greater than the sum of its parts. And the specific pattern of the ties is crucial to understanding how MOOCs function. It is apt to adopt a methodology influenced by social network analysis for dynamic team formation in MOOCs, because it is ultimately the students who choose the structure of their ego network in the MOOC. Firstly, students informally decide how many fellow mates they are connected to. For example, students may require one classmate for clearing a technical course related issue, while they would want to engage with a handful of students for discussing an essay. Secondly, students influence how densely interconnected their peer group in the MOOC is. For example, some students might start threads or subthreads of discussion on interesting course topics, so that all their virtual friends can meet each other. Thirdly, students control how central they are to the network. For example, some students may be the life of the MOOC, mingling with most discussion initiators and responders, while others may prefer to stay on the sidelines and just observe discussion forum activities with minimal participation. Considering the conversation networks in MOOCs might be sparse, based on the 90-9-1 rule in online communities, one limitation of the above proposed approach might be the fact that it would be hard or unfair to refine the students' skills, because knowledgeable students might not join the conversations at all, yet students who are interacting with many threads might just complain that they encounter the same problem and not be able to solve it. Thus, it would be intriguing to think about incorporating another dimension in this MOOC network, where the value of each post or reply is also under consideration. Overall, the proposal aimed at extracting value from the MOOC social network, and using it for forming teams automatically. This will not only aid course instructors, but also guide students to improve their participation and collaboration in the MOOC. Providing students with fruitful ways on engagement in the MOOC is a first step towards preventing huge attrition rates that have plagued the first generation of MOOCs.

## ACKNOWLEDGMENT

The author would like to thank Mr. David Adamson, Phd candidate at Language Technologies Institute (LTI), Carnegie Mellon University, for preparation of the Coursera MOOC dataset.